# Liquidus temperature nonlinear modeling of silicates $SiO_2$-$R_2O$-RO

Patrick dos Anjos[1]*

*Federal Institute of Espírito Santo (IFES), Vitória, ES, Brazil*

## Abstract

The liquidus temperature is an important parameter in understanding the crystalline behavior of materials and in the operation of blast furnaces. Its modeling can be carried out by linear and nonlinear methods through data, considering the artificial neural network a modeling method with high efficiency because it presents the theorem of universal approximation and with that better performances and possibility of greater oscillations. The best linear model and the best nonlinear model were modeled by structural parameters and presented a good numerical approximation, thus demonstrating that mathematical modeling can be performed using structural arguments and also showing a dimensionality reduction method for modeling a thermophysical property of the materials.

## 1  Introduction

The liquidus temperature ($T_{liq}$) is one of the most important thermophysical properties of blast furnace slag [1], with extreme importance in its operation [2], and is also an important physical parameter for understanding the crystalline behavior of slag [3]. $T_{liq}$ can be defined as the temperature at which a material becomes completely liquid, differing from the solidification temperature ($T_{sol}$) (or break temperature ($T_{br}$), when related to viscosity) due to the phenomenon of super-cooling [4].

The transition temperature ($T_{tt}$) is the temperature at which the thermal conductivity ($\lambda$) has a physical behavior change in the heating of slag [6]. At temperatures above $T_{tt}$ ($T>T_{tt}$) the $\lambda xT$ relationship has an exponential relationship and when the temperature decreases the thermal conductivity tends to increase, probably caused by the difference in the precipitation rate of the crystal together with the effects of the glassy and crystalline state. When $T<T_{tt}$ there is a linear relationship in the $\lambda xT$ relationship, where the conductivity decreases with decreasing temperature [5]. Li et al. (2021) mathematically modeled the $T_{tt}$ of $SiO_2$-$Al_2O_3$-CaO-$Fe_2O_3$-MgO slags in relation to $T_{liq}$ (Equation 1).

$$T_{tt} = 1.067 T_{liq} + 243.2 \qquad (1)$$

---

[1]  E-mail: patrick.dosanjos@outlook.com

$T_{liq}$ is directly related to viscosity. Viscosity can be modeled using the Arrhenius equation (Equation 2), where *A* is constant, *R* is the universal gas constant, *T* is temperature, *Q* is activation energy and *η* is viscosity.

$$\eta = A \exp\left(\frac{B}{QT}\right) \qquad (2)$$

The activation energy can be defined as the first derivative of the viscosity logarithm with respect to 1/T, obtaining Equation 3.

$$Q = R \frac{\partial \log \eta}{\partial \left(\frac{1}{T}\right)} \qquad (3)$$

In temperatures above $T_{liq}$, applied to the second derivative of Equation 2 in relation to temperature $((\partial^2 Q/\partial T^2)/R)$, the activation energy Q has a constant value [6] (Figure 1).

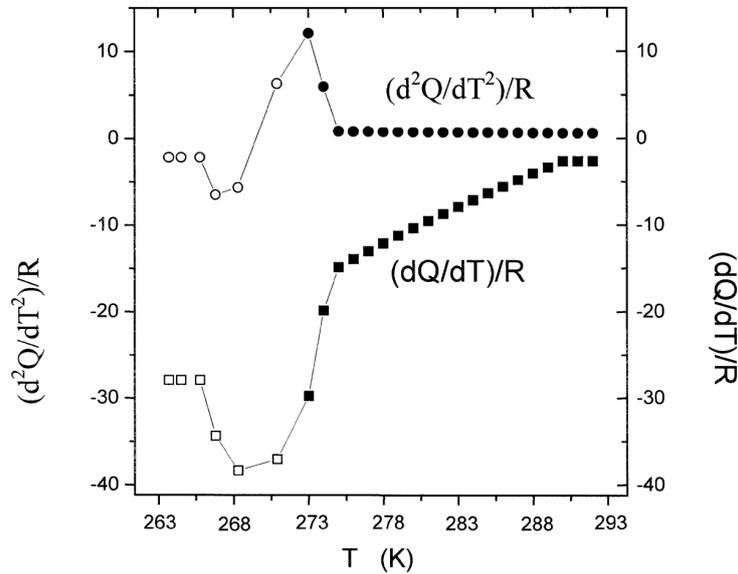

Figure 1 — The first $((dQ/dT)/R)$ and second $((d^2Q/dT^2)/R)$ derivative of activation energy *Q* in relation to Arrhenius equation [6].

The structure of a chemical system is related to properties such as electrical conductivity [7], thermal conductivity [8], hardness [9], Poisson coefficient [10] and viscosity [4]. The structure can be defined by the NBO/T depolymerization parameter, which indicates the number of non-bridging oxygen (NBO) by tetrahedron (T). In $SiO_2$-$R_2O$-RO systems the NBO/T parameter is indicated by Equation 4 [4,11,12], where RO=CaO, MgO, …, $R_2O$=$Li_2O$, $Na_2O$, …, and X the mole fraction.

$$NBO/T = \frac{2\left(\sum XRO + \sum XR_2O\right)}{XSiO_2} \qquad (4)$$

Tetrahedrons are structures composed of a central ion surrounded by oxygen atoms, composing different structural parameters. The NBO/T can have value 0, 1, 2, 3 and 4, corresponding to the structural parameters $Si_2O_8^{8-}$, $Si_2O_7^{6-}$, $Si_2O_6^{4-}$, $Si_2O_5^{2-}$ and $Si_2O_4$ respectively. These structural parameters have varied reactions, as shown in Equation 5, Equation 6 and Equation 7 [13].

$$2Si_2O_7^{6-} = Si_2O_8^{8-} + Si_2O_6^{4-} \quad (5)$$

$$3Si_2O_6^{4-} = Si_2O_8^{8-} + Si_2O_5^{2-} \quad (6)$$

$$2Si_2O_5^{2-} = Si_2O_6^{4-} + Si_2O_4 \quad (7)$$

The formation of NBO results from the addition of modifying species such as $R^+$ and $R^{2+}$ ions in the silicate structure by dissociation of RO and $R_2O$ oxides. The action of $R^+$ ions is indicated by Equation 8, where there is a weakening of the structure and the $R^{2+}$ ions act by breaking the bonds between the tetrahedrons [13] (Equation 9).

$$\equiv Si\text{-}O\text{-}Si\equiv\; +\; R_2O \rightarrow\; \equiv Si\text{-}O\text{-}R\; +\; R\text{-}O\text{-}Si\equiv \quad (8)$$

$$\equiv Si\text{-}O\text{-}Si\equiv\; +\; RO \rightarrow\; \equiv Si\text{-}O\text{-}R\text{-}O\text{-}Si\equiv \quad (9)$$

As the viscosity influences the Tliq and the structure influences the viscosity, consequently the structure influences the $T_{liq}$ by hypothetical syllogism [14]. $T_{liq}$ can be measured experimentally through experiments intermediated by differential scanning calorimetry (DSC) [3], and can be mathematically modeled by linear modeling [3,4] and by nonlinear modeling [1,2] in relation to variables such as chemical composition. Linear modeling can be conducted using the method of least squares with the possibility of regularization l1 or l2 [15] and nonlinear modeling can be performed using the method of nonlinear least squares [16] and the use of artificial neural networks (ANN) [17] that present advantages over other mathematical models for having complex interactions between variables [18].

The aim of this work is to mathematically model the $T_{liq}$ of $SiO_2$-$R_2O$-RO silicates by linear modeling using the least squares method with l1 regularization and by nonlinear modeling using artificial neural networks.

## 2 Materials and Methods

### 2.1 Database

The SciGlass [19] database was used to provide data on chemical composition (%mol) and liquidus temperature ($T_{liq}$), also used to predict the properties of chalcogenic glasses [20], refractive index of optical materials [21] and the viscosity of oxides [22]. The chemical system $SiO_2$-CaO-$K_2O$-$Na_2O$-PbO-$Li_2O$-MgO-SrO-BaO-ZnO was selected with the $T_{liq}$ related to each chemical composition of the database. A preprocessing of the database was carried out to relate the NBO/T parameter to the removal of outliers considering Equation 10.

$$\mu_{NBO/T} - 3\sigma_{NBO/T} \leq NBO/T \leq \mu_{NBO/T} + 3\sigma_{NBO/T} \quad (10)$$

$\mu_{NBO/T}$ is the mean of the NBO/T and $\sigma_{NBO/T}$ equals the standard deviation of the NBO/T in the database. There is the polymerization parameter $Q^n$ where n is equivalent to the bridging oxygen (BO) number (BO = NBO – 4) [13]. Several $Q^n$ values coexist in the same chemical system [23-28]. The NBO/T parameter is a continuous variable [29] and its distribution in the database can be seen in Figure 2.

The mathematical modeling of $T_{liq}$ based on the chemical composition [1-4] is a frequent method and there are no extensive studies on this mathematical modeling based on the structural parameters that make up the NBO/T parameter. The database was related with the parameters $\Sigma XRO$, $\Sigma XR_2O$ and $XSiO_2$ to be the independent variables or predictors of the dependent or predicted variable $T_{liq}$. The database was normalized between 0 and 1. Normalization aims the process of training and validation of the mathematical modeling by maintaining the original distribution of the variables, unlike standardization, which changes the data to have an arithmetic mean and standard deviation equal to 0 and 1, respectively, with normal distribution [22]. The $T_{liq}$ data in the database and the normalized $T_{liq}$ data can be seen in Figure 3. Training data, validation data and test data were taken from the database. Validation data and test data were **not** used in training in linear modeling and non-linear modeling.

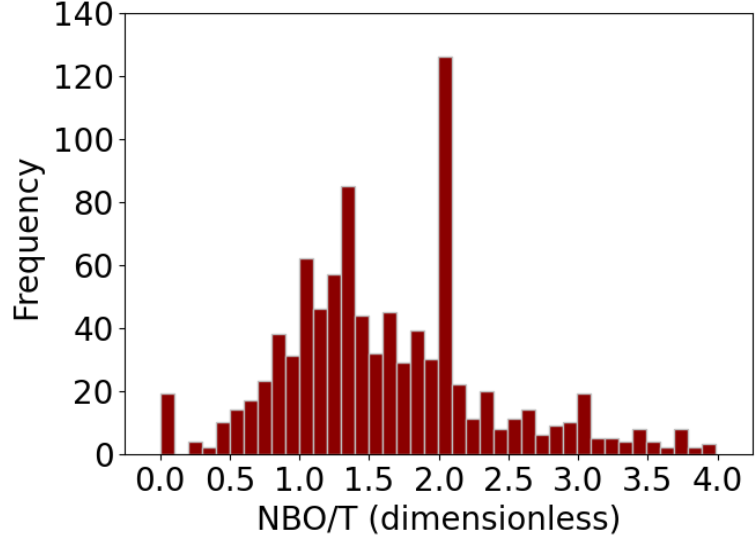

Figure 2. NBO/T distribution in database

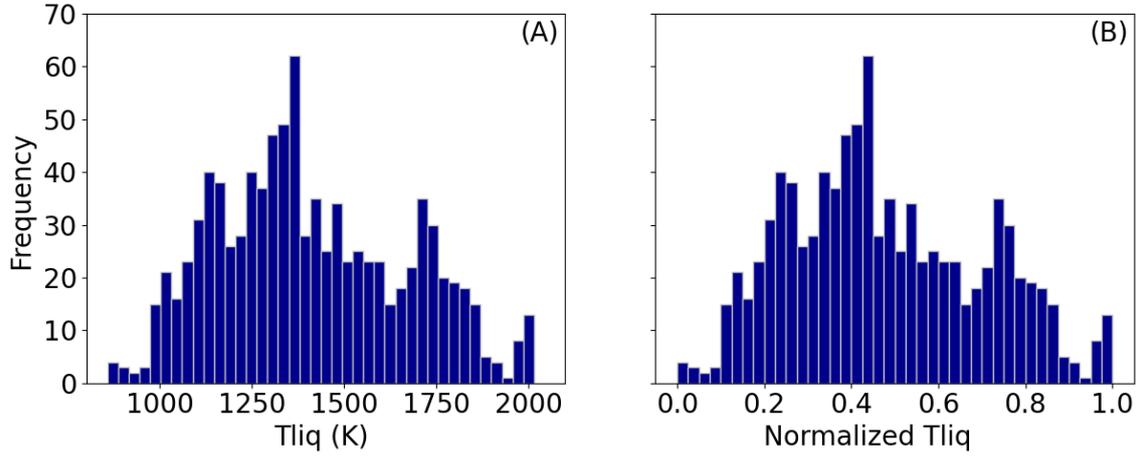

Figure 3. T$_{liq}$ distribution in database (A) with raw data and (B) with normalized database

## 2.2 Linear Modeling

Linear modeling was carried out by linear regression using the least squares method with regularization l1 (Equation 11) as it presents feature selection and non-linearity between the regularization parameter α and the vector of independent variables [15].

$$\text{minimize } \| Ax - y \|_2^2 + \alpha \| x \|_1 \tag{11}$$

*A* is the vector of constants related to the predictor variables *x*, *y* the predicted variable and *α* the regularization parameter l1.
When the α parameter has the value 0, it indicates that there is no regularization l1 and the linear modeling is governed by the ordinary least squares method. Numerical analyzes were performed by varying the α parameter to present the lowest mean absolute error (MAE) (Equation 12) and the lowest variance (S2) (Equation 13) between the data predicted by the built model and the test database.

$$MAE = \frac{\sum |y_{true} - y_{pred.}|}{N} \tag{12}$$

$$S^2 = \frac{\sum (dev_i - \bar{x}_{dev.})^2}{N-1} \tag{13}$$

$N$ is the amount of data, $y_{true}$ is the $T_{liq}$ of the test database, $y_{pred.}$ is the $T_{liq}$ predicted by the model, $dev_i$ the i-th deviation considering deviation the difference between $y_{true}$ and $y_{pred.}$ and $\bar{x}_{dev.}$ the arithmetic mean of the deviation.

**2.3 Nonlinear Modeling**

The nonlinear modeling was carried out using artificial neural networks through Adam [30] optimization with the initialization of weights and biases through the uniform Glorot distribution [31]. In the training of artificial neural networks, the error was computed through the mean squared error (MSE) (Equation 14) and the EarlyStopping [32] technique was used to avoid underfitting or overfitting.

$$MSE = \frac{\sum (y_{true} - y_{pred.})^2}{N} \tag{14}$$

The universal approximation theory states that an artificial neural network with a hidden layer (depth-1) can approximate any continuous function [33]. But depth-1 artificial neural networks have few oscillations [34] and the theory of universal approximation does not discuss Rademacher Complexity [35-36] or the minimum width required for universal approximation [37]. There is depth approximation in ReLU artificial neural networks with depth $d \leq k+2$ and width $w \leq k^{3/2}$ with $k \geq n+4$ and n consisting of the number of elements in the vector of predictor variables [38]. Artificial neural networks were constructed with the approximation by depth using a $k=n+4$ for the nonlinear models to have greater oscillations [34] and better performances [39].

**2.3 Statistical Evaluation**

The mean absolute error values (Equation 12) and variance (Equation 13) of the deviations of each linear model and nonlinear model were used, and a parameter was considered that relates these values, represented by Equation 14.

$$M_2 = \frac{MAE + \sqrt{S^2}}{2} \tag{14}$$

The $M_2$ parameter relates the first moment and the second central moment [40] and compares the error and variability of the linear and nonlinear models constructed. The best linear model and the best nonlinear model are considered those with the lowest $M_2$ value because they have the lowest error and lowest joint variability. The Mean Absolute Percent Error (MAPE) metrics (Equation 15) were also used for the best linear model and for the best nonlinear model.

$$MAPE = \frac{1}{N} \frac{\sum (y_{true} - y_{pred.})}{y_{true}} \tag{15}$$

Sensitivity analysis is a measure of the importance of predictor variables in the predicted variable and presents the significance of predictor variables in artificial neural networks. The sensitivity analysis was carried out in the non-linear model with the best performance by the W coefficient (Equation 16) which presents the relationship between the training error with all predictive variables ($Error_t$) with the error of the re-implementation of the training with the withdrawal of a variable i among the predictor variables ($Error_{t-i}$) [41].

$$W = \frac{Error_{t-1}}{Error_t} \tag{16}$$

When the coefficient W results in 1 or less (W≤1) it indicates that the removal of the variable i among the predictor variables has no importance in the artificial neural network. The coefficient W resulting in a number greater than 1 (W>1) denotes that the removal of the variable i among the predictor variables is important in the artificial neural network [41]. W coefficients were computed with the removal of the variables $\Sigma XRO$, $\Sigma XR_2O$ and $XSiO_2$ to present the importance of each parameter in the best artificial neural network.

## 3 Results and discussion

### 3.1 Linear Modeling

Linear modeling performed by linear regression with l1 regularization with variation of the regularization parameter $\alpha$ between -100 and 100, obtaining an optimal value of 0. That is, even with the use of the l1 regularization technique, the best linear model was established by optimization by the method of ordinary least squares. Equation 17 demonstrates the linear regression between the structural parameters $\Sigma XRO$, $\Sigma XR_2O$ and $XSiO_2$ with respect to liquidus temperature with the training data.

$$T_{liq} (K) = 0.315 \Sigma XRO - 0.102\ \Sigma XR_2O + 0.244 XSiO_2 + 0.319 \qquad (14)$$

In the test data, the linear model represented by Equation 17 presented MAE of 154.13K and a variance of 14685.99$K^2$, with M2 of 137.66K an MAPE of 11.54% (Table 1). The difference between the predicted test data and the database test data can be seen in Figure 4.

Table 1. Statistical Evaluation of Linear Modeling of $T_{liq}$

| Model | MAE (K) | Variance ($K^2$) | $M_2$ (K) | MAPE (%) |
|---|---|---|---|---|
| Linear Modeling | 154.13 | 14685.99 | 137.66 | 11.54 |

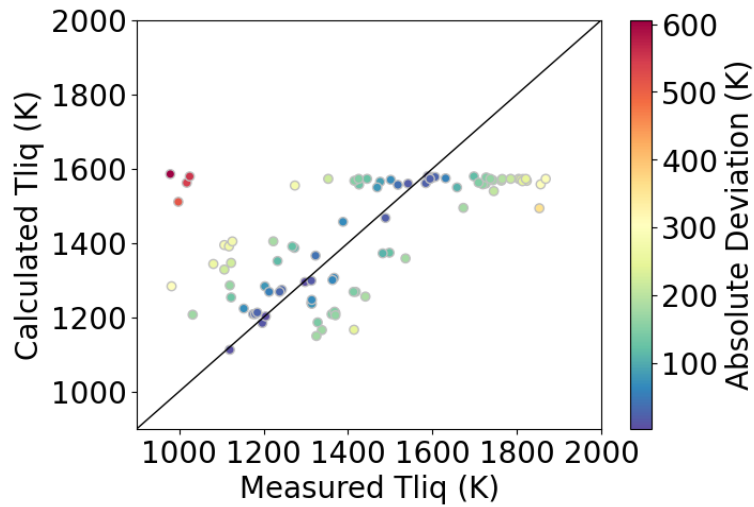

Figure 4. Test data (Measured $T_{liq}$) related to predicted data (Calculated $T_{liq}$) resulted to linear modeling (K)

### 3.2 Nonlinear Modeling

The constructed artificial neural networks add up to 144 different nonlinear models. The test results of the artificial neural networks with the test database, in relation to the MAE, can be seen in Figure 5.

With the MAE metric, the best artificial neural network had a width of 18 and a depth of 1 (1-18) with a value of 129.28K (Table 2). The artificial neural network with width 2 and depth 17 presented a MAE value equal to 129.31K and the one with width 6 and depth 4 resulted in a MAE equal to 129.91K, with values very close to the best artificial neural network with the smallest MAE presented in the test data.

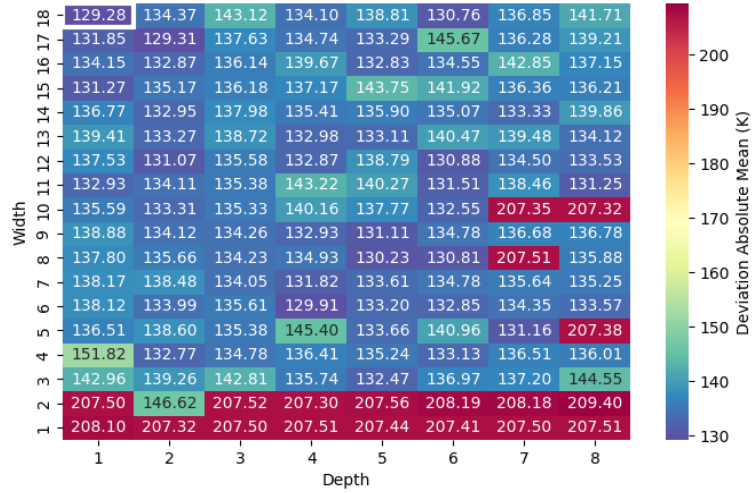

Figure 5. MAE heatmap in nonlinear modeling (highlighted artificial neural network 1-depth 18-width)

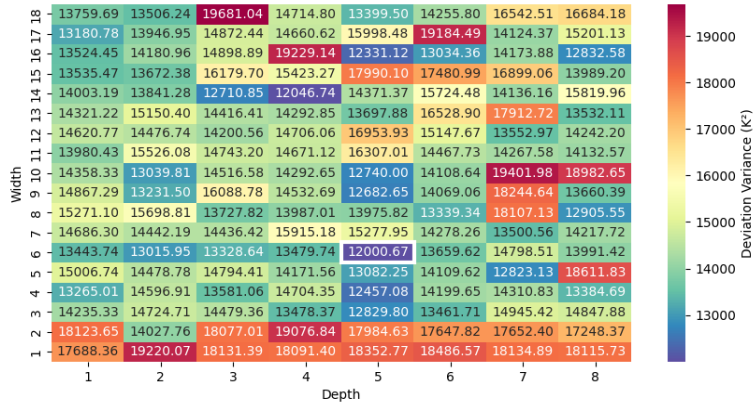

Figure 6. Variance heatmap in nonlinear modeling (highlighted artificial neural network 5-depth 6-width)

At variance, the artificial neural networks demonstrated behavior as described in Figure 6. The best artificial neural network presented the variance value equal to 12000.67$K^2$ with a width of 6 and depth 5 (5-6) (Table 2). The artificial neural network with width 14 and depth 4 presented a variance of 12046.74$K^2$ with an approximate value of the best artificial neural network by the variance metric in the test data. The calculation of the $M_2$ of each trained artificial neural network was established by Equation 14 and can be seen in Figure 7.

The best artificial neural network by the parameter $M_2$ presented a width of 6 and depth 5 (5-6) with a value of 121.37K (Table 2), considered the best nonlinear model. By the $M_2$ parameter, the artificial neural networks with width 9 and 16 with depth 5 presented values of 121.86K and 121.94K respectively, representing approximate values in relation to the best artificial neural network by the $M_2$ parameter in relation to the test data. The best nonlinear model considered is the artificial neural network with depth 5 and width 6 presented the train and validation curve according to Figure 8. The artificial neural network 5-6 presented a MAPE of 9.92%, superior to the best linear model, thus denoting that they have nonlinear interactions between the variables $\Sigma XRO$, $\Sigma XR_2O$ and $XSiO_2$ and the liquidus temperature (Table 2).

In Figure 7, it can be seen that the number of epochs for training the artificial neural network 5-6 exceeds the value of 10000 and does not present a maximum multiple of 10, for example, due to the action of the EarlyStopping technique. With the test data, the relationship was established between the values predicted by the artificial neural network 5-6 with the values from the test database, shown in Figure 9.

Table 2. Statistical Evaluation of Nonlinear Modeling of $T_{liq}$

| Model | MAE (K) | Variance (K$^2$) | M$_2$ (K) | MAPE (%) |
|---|---|---|---|---|
| Nonlinear Modeling | 129.28 | 12000.67 | 121.37 | 9.92 |

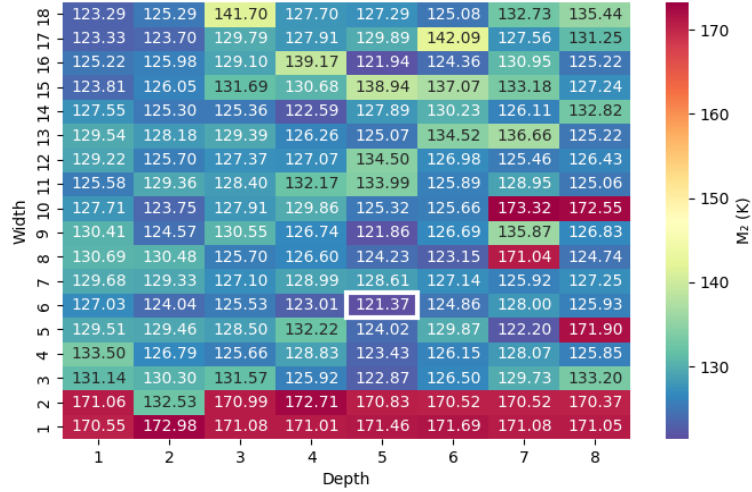

Figure 7. M$_2$ heatmap in nonlinear modeling (highlighted artificial neural network 5-depth 6-width)

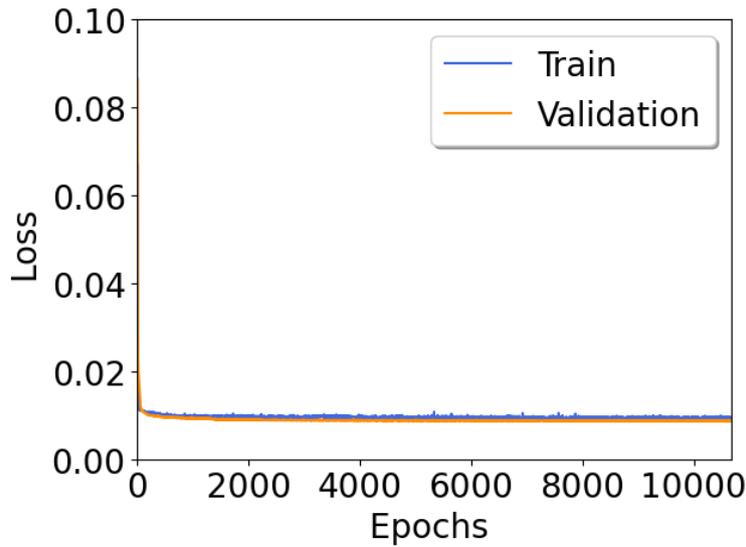

Figure 8. Loss versus Epochs in artificial neural network 5-6

In relation to the sensitivity analysis by the coefficient W, the variables ΣXRO, ΣXR2O and $XSiO_2$ presented values of 1.1977, 1.1202 and 1.1507 respectively (Table 3). Thus, all the variables used as arguments in the best nonlinear modeling, considering the artificial neural network 5-6, demonstrated importance in the prediction of the liquidus temperature, with the parameter ΣXRO considered the most important variable (1.1977), ΣXR$_2$O with the lowest weight (1.1202) $XSiO_2$ with intermediate weight (1.1507).

In terms of relative importance (%), the variable ΣXRO represents 34.53%, $XSiO_2$ with 33.17% and ΣXR$_2$O with 32.30%, considered variables with weights of similar importance by approximation (Table 3).

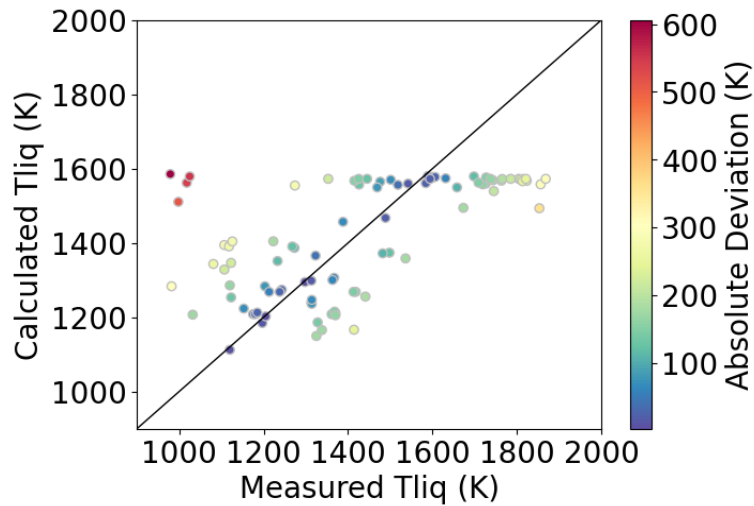

Figure 9. Test data (Measured $T_{liq}$) related to predicted data (Calculated $T_{liq}$) resulted to nonlinear modeling (K)

Table 3. Statistical Evaluation of Nonlinear Modeling of $T_{liq}$

| Sensitvity Analysis | $\Sigma XRO$ | $\Sigma XR_2O$ | $XSiO_2$ |
|---|---|---|---|
| Coefficient W (Relative Importance) | 1.1977 (34.53%) | 1.1202 (32.30%) | 1.1507 (33.17%) |

# 4   Conclusions

Linear modeling and nonlinear modeling of $SiO_2$-CaO-$K_2O$-$Na_2O$-PbO-$Li_2O$-MgO-SrO-BaO-ZnO chemical systems using the structural parameters $\Sigma XRO$, $\Sigma XR_2O$ and $XSiO_2$. They presented good results in relation to mathematical modeling. The linear modeling with greater efficiency presented a MAE of 154.13K and a variance of 14685.99$K^2$, with $M_2$ of 137.66K an MAPE of 11.54%. The most efficient nonlinear modeling, represented by an artificial neural network with width 6 and depth 5 (5-6), presented a MAE of 133.20K, a variance of 12000.67$K^2$, $M_2$ of 121.37 and a MAPE of 9.92%.

With this, it is possible to denote an approximation with the use of mathematical modeling, using linear modeling and nonlinear modeling, using the structural parameters of the NBO/T. There is also a dimensionality reduction in the mathematical modeling with the reduction of the $SiO_2$-CaO-$K_2O$-$Na_2O$-PbO-$Li_2O$-MgO-SrO-BaO-ZnO chemical system for the parameters $\Sigma XRO$, $\Sigma XR_2O$ and $XSiO_2$.